\documentclass[journal]{IEEEtran}
\usepackage{amsmath}
\usepackage{amssymb}
\usepackage{graphicx}
\usepackage{makecell}
\usepackage{multirow}
\usepackage{subcaption}
\usepackage{xcolor}
\usepackage{cellspace}
\usepackage{stackengine}
\setstackEOL{\\}

\DeclareMathOperator*{\argmax}{arg\,max}

\title{Understanding the Perceived Quality of Video Predictions}
\author{Nagabhushan~Somraj,
        Manoj~Surya~Kashi,
        S. P. Arun and
        Rajiv Soundararajan%
\thanks{N. Somraj, M. S. Kashi and R. Soundararajan are with the Department of Electrical Communication Engineering, Indian Institute of Science, Bengaluru 560012, India (email: nagabhushans@iisc.ac.in; manojsk@iisc.ac.in; rajivs@iisc.ac.in). S. P. Arun is with the Centre for Neuroscience, Indian Institute of Science, Bengaluru 560012, India (email: sparun@iisc.ac.in). This research was supported in part by Pratiksha Trust grants (FG/PTCH-19-2068 and FG/PTCH-20-2068). The first author was supported by the Prime Minister's Research Fellowship awarded by the Ministry of Human Resources Development, Government of India.}
}

\begin{document}
    \maketitle
    \begin{abstract}
        The study of video prediction models is believed to be a fundamental approach to representation learning for videos.
        While a plethora of generative models for predicting the future frame pixel values given the past few frames exist, the quantitative evaluation of the predicted frames has been found to be extremely challenging.
        In this context, we study the problem of quality assessment of predicted videos.
        We create the Indian Institute of Science Predicted Videos Quality Assessment (IISc PVQA) Database consisting of 300 videos, obtained by applying different prediction models on different datasets, and accompanying human opinion scores.
        We collected subjective ratings of quality from 50 human participants for these videos.
        Our subjective study reveals that human observers were highly consistent in their judgments of quality of predicted videos.
        We benchmark several popularly used measures for evaluating video prediction and show that they do not adequately correlate with these subjective scores.
        We introduce two new features to effectively capture the quality of predicted videos, motion-compensated cosine similarities of deep features of predicted frames with past frames, and deep features extracted from rescaled frame differences.
        We show that our feature design leads to state-of-the-art quality prediction in accordance with human judgments on our IISc PVQA Database.
        The database and code are publicly available on our project website: https://nagabhushansn95.github.io/publications/2020/pvqa
    \end{abstract}

    \IEEEpeerreviewmaketitle

    \section{Introduction}\label{sec:introduction}
    Video prediction refers to the problem of generating pixels of future frames given context information in the form of past frames.
    The problem has attracted a lot of attention in the context of generative video models.
    The ability to predict the future accurately has applications in various domains, including robotics for path planning, self-driving cars, anomaly detection~\cite{liu2018future}, and video compression.
    It is also shown that solving this problem offers a fundamental approach to learning internal representations of videos~\cite{srivastava2015unsupervised, byeon2018contextvp}.
    Further, the problem also helps in understanding interactions of physical objects in the real world~\cite{finn2016unsupervised, janner2018reasoning}.
    Although, there may be applications where we are only interested in predicting features in future frames~\cite{agrawal2016learning}, predicting all pixels in future frames ~\cite{srivastava2015unsupervised,mathieu2016deep} allows for rich self-supervision, a visual interpretation of the predicted frames, and a more generic approach to learning across different applications.
    The video prediction problem leads to an important question of how to generically evaluate the quality of the predicted videos in a task-free viewing condition.

    While there exists a rich body of work on video prediction using generative models, the design of methods for evaluating the quality of the videos has received much less attention.
    Simple signal fidelity measures such as mean squared error (MSE) or the structural similarity (SSIM) index~\cite{wang2004image} can be computed in scenarios where a reference future video sequence is available.
    However, for a given context, there might exist a multitude of possible future video trajectories that are natural looking.
    It would be unfair to compare such predicted videos against a given future realization.

    The quality assessment of predicted videos needs to capture multiple notions.
    Indeed, video prediction researchers have identified the sharpness of predicted frames as an important evaluation tool~\cite{mathieu2016deep}.
    The spatial quality of predicted video frames is also influenced by the realism of object shapes and texture.
    Object motion and temporal consistency are important elements of the quality of predicted videos.
    Further, the events unfolding in a video need to make logical sense.
    Thus, the quality assessment of predicted videos appears to involve elements of both early and later stages of human vision systems.
    In this work, we particularly focus on predicted videos obtained using generative prediction models given the rich literature on this topic.

    The quality assessment of predicted videos presents its own challenges when compared to generic video quality assessment (VQA) as described above. However, the problem formulation also allows one to exploit more information when compared to VQA without using a reference video. In particular, video prediction methods are typically applied given a few context frames which are of high quality. Thus, the quality of predicted videos could be assessed by exploiting information available in the context frames. We believe that assessment of aspects such as object shapes, their evolution, texture and blur in the predicted frames can be achieved more reliably by making use of such information. This makes the problem of assessing the quality of predicted videos quite different from that of classical VQA.

    \begin{figure*}
        \includegraphics[width=\linewidth]{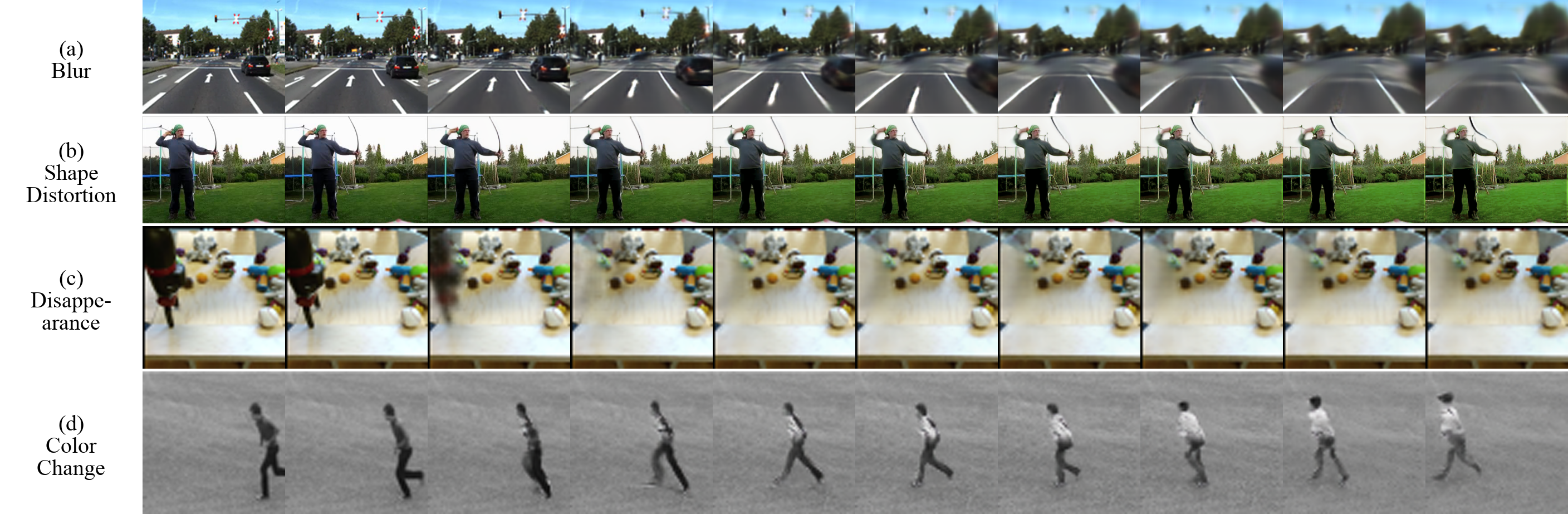}
        \caption{Example distortions observed in video frames in our database.
        The sequence of images in each row corresponds to the frames of a video.
        Starting from the first frame, we show every second frame.
        The first two frames correspond to the context frames, and the next eight frames correspond to the predicted frames.
        The mean opinion score (MOS) obtained from the subjective study is also shown below for each video.
            (a) In the first video, we observe the gradual increase in blur with deeper prediction in time.
            MOS: 40.01.
            (b) In the video in the second row, the shape of the bow gets distorted over time.
            While blur is global, shape distortion is highly localized.
            MOS: 45.41.
            (c) The video in the third row shows the disappearance of the robotic arm.
            MOS: 41.01.
            (d) In the video in the fourth row, as the person runs from right towards left, the color of his shirt changes from white to black.
            MOS: 55.24.
            The videos can be viewed on our project website.
        }
        \label{fig:diff-distortions}
    \end{figure*}

    The main focus of our work is in the subjective and objective study of the quality of predicted videos.
    Recently, small scale subjective studies based on pairwise comparisons have been carried out to prove the effectiveness of specific video prediction models~\cite{lee2018stochastic}.
    While human opinion might be the best subjective measure of quality, collecting such human data is cumbersome, and it is desirable to have an objective, automatic measure of quality that can be evaluated on any video.
    We believe that a continuous valued measure that can be evaluated on any predicted video will be useful in comparing various prediction methods.

    Very recently, the Fr\'echet video distance (FVD) was introduced to evaluate generative models and validated using a subjective study~\cite{unterthiner2019fvd}.
    The distance is meant to be applied on a collection of generated videos instead of individual videos and is thus different from our goal to measure quality.
    Further, the study is designed primarily to prove the effectiveness of FVD while we seek to design a study that can help benchmark and advance research in measuring the quality of any predicted video.
    To the best of our knowledge, there exists no human study on predicted videos that measures quality of individual videos on a continuous scale.

    \subsection{Overview of Contributions}\label{subsec:contributions-overview}
    Our main contributions in this work are in the creation of a database of predicted videos, design of a subjective study, benchmarking of existing objective methods used to evaluate quality, and introduction of mechanisms leading to improved prediction of video quality.
    We create the Indian Institute of Science Predicted Videos Quality Assessment (IISc PVQA) Database consisting of 300 videos, each consisting of 20 frames, obtained from a variety of different prediction models.
    The videos are generated by applying video prediction models on databases typically used to evaluate them.
    Our database contains a variety of sources of distortions such as blurred frames, frames with distorted object shapes, temporal color variations, and sudden appearance or disappearance of objects, as shown in Fig.~\ref{fig:diff-distortions}.
    Thus our database is diverse in terms of content and distortions.

    We conduct a subjective study involving 50 human subjects resulting in a total of 6000 video ratings under calibrated conditions.
    Since the videos from different databases are available at different resolutions and might bias the quality scores, we adopt a double stimulus continuous quality assessment method.
    In our study, a pair of videos is shown, one being the test video and the other, a different natural video with similar content.

    We benchmark several popular measures to evaluate predicted videos, such as MSE, SSIM and deep network based loss functions, against the subjective scores of video quality.
    We show that these measures do not correlate well with the subjective scores since they evaluate the predicted videos by assuming a fixed trajectory of the reference.
    We also show that popular no-reference video QA algorithms do not match well with subjective judgements of video quality since they are not designed to capture the artifacts that arise in video prediction.

    Finally, we introduce two novel sets of features to effectively predict the quality of predicted videos.
    The first set of features is based on computing cosine similarities of deep features of past frames with corresponding motion-compensated features from the predicted frames.
    This helps capture object blur, shape, and color distortions in a robust fashion by comparing with the past frames.
    Secondly, we rescale frame differences of adjacent frames of the predicted video to appear like an image and extract corresponding deep features to capture object shape variations in regions containing motion.
    We show that these features can effectively predict video quality by achieving the state-of-the-art performance in terms of correlation with the subjective scores.

    We summarize the main contributions of our work as follows:
    \begin{enumerate}
        \item We introduce the IISc PVQA database of 300 videos predicted using a variety of models and based on multiple datasets.
        \item We conduct a behavioral study with 50 subjects to measure the quality of the predicted videos through a double stimulus scoring mechanism.
        \item We benchmark several metrics popularly used in video prediction evaluation and show that existing metrics correlate poorly with human perception of video quality.
        \item We propose novel features based on motion-compensated cosine similarities and rescaled frame differences and show that they are useful in predicting quality in a manner that agrees very well with human perception.
    \end{enumerate}

    The rest of the paper is organized as follows.
    In Section~\ref{sec:related-work}, we survey related work.
    We describe the predicted video quality assessment database and the subjective study in Section~\ref{sec:database}.
    We introduce our quality assessment features in Section~\ref{sec:deep-feature-processing}.
    We present detailed experiments and ablation studies in Section~\ref{sec:experiments} and finally conclude the paper in Section~\ref{sec:conclusion}.

    \section{Related Work}\label{sec:related-work}

    \subsection{Video Prediction}\label{subsec:related-work-video-prediction}
    Predicting video frames from past frames through classical methods and deep learning has been an important aspect of video compression~\cite{wiegand2003overview,choi2020deep} for long.
    However, a lot of progress has been made in video prediction with the advent of deep image generation models ~\cite{goodfellow2014gan}.
    One of the primary reasons for advancement in video prediction has been due to the use of adversarial loss functions~\cite{mathieu2016deep,villegas2017mcnet}.
    Various researchers have approached video prediction through decomposition of motion and content~\cite{villegas2017mcnet}, motion modelling through filter kernels~\cite{finn2016unsupervised}, predictive coding~\cite{lotter2017deep} and 3D Long Short Term Memory (3D LSTM)~\cite{wang2018eidetic}.
    Researchers have also developed stochastic generation models~\cite{denton2018stochastic,villegas2019high} to account for the uncertainty in prediction and the possibility of a multitude of future trajectories given the context frames.
    A detailed survey of deep learning based methods in video prediction can be found in~\cite{oprea2020review}.

    \subsection{Evaluation methods for video prediction models}\label{subsec:related-work-eval-methods}
    The most popular method of evaluating predicted video frames is using MSE or the SSIM index~\cite{wang2004image}.
    In a variant of MSE, areas with higher motion are weighted preferentially using optical flow based weights~\cite{mathieu2016deep}.
    Other measures that involve comparison with a reference include squared error~\cite{janner2018reasoning} and cosine similarity~\cite{lee2018stochastic,kumar2019videoflow} in the pre-trained VGG net~\cite{simonyan2015very} feature space.
    The inception score for images~\cite{salimans2016improved} has also been applied to evaluate generated video frames~\cite{he2018probabilistic,xu2018video}.
    The image inception distance has been extended to videos through FVD~\cite{unterthiner2019fvd}.
    In particular, features based on Inflated 3D Convnet are used to compute a distance measure between a set of generated videos and a database of pristine videos.
    FVD was validated using a human study through pairwise tests on the BAIR dataset~\cite{ebert2017self}.
    Further 2AFC experiments were conducted to evaluate few video prediction models~\cite{lee2018stochastic, wichers2018hierarchical}.

    \subsection{Video quality assessment}\label{subsec:related-work-vqa}
    Video quality assessment (VQA) has been studied quite extensively over the last decade or so with the conduct of several studies of subjective quality and the design of successful objective algorithms.
    Publicly available VQA databases include those containing synthetic distortions such as the LIVE VQA database~\cite{seshadrinathan2010study} or those containing authentic camera captured distortions such as the LIVE Video Quality Challenge (LIVE VQC) Database~\cite{sinno2019large} and the KoNViD-1k database~\cite{hosu2017konstanz}.
    VQA algorithms are broadly divided into three categories, full reference (FR), reduced reference (RR) and no reference algorithms (NR).
    FR VQA algorithms utilize a reference video to predict the quality of a distorted video by exploiting both spatial and temporal similarity.
    RR VQA algorithms extract quality aware features from reference video and compare them with features of distorted video to predict its quality.
    Some examples of successful FR and RR VQA algorithms that exploit spatio-temporal information include MOVIE~\cite{seshadrinathan2009motion}, ST-MAD~\cite{vu2011spatiotemporal}, ST-RRED~\cite{soundararajan2013video} and VMAF~\cite{li2018vmaf}.
    These algorithms operate either by computing spatio-temporal transformations or obtain quality features separately in the spatial and temporal domains and combine them.

    The lack of availability of a true reference in several scenarios motivates the design of NR algorithms.
    The NR VQA problem has been found to be much more challenging than the FR problem, and current NR algorithms are not yet as successful as the FR algorithms.
    Video BLIINDS~\cite{saad2014blind}, SACONVA~\cite{li2016no} and TLVQM~\cite{korhonen2019two} are examples of methods that have been able to approach the performance of FR algorithms.
    Recently, deep neural networks have been used to obtain good performance on authentic distortions~\cite{li2019quality}.
    Nevertheless, the use of convolutional neural networks to design successful NR VQA algorithms is still a nascent and active area of research.

    \section{Predicted Videos Quality Assessment Database}\label{sec:database}
    We now describe in detail, the IISc Predicted Videos Quality Assessment (IISc PVQA) database, our subjective study and important observations from the study.

    \subsection{Database}\label{subsec:database}
    The videos in our database are generated by various video prediction algorithms.
    These video prediction algorithms are trained on a variety of datasets containing human actions, sports videos, vehicle driving, and robot pushing videos.
    In our database, we use a combination of publicly available pre-trained models of different prediction algorithms and also models that we train on other datasets.

    \textit{Datasets}:
    \begin{table*}
        \begin{center}
            \caption{Number of videos from different datasets}
            \smallskip
            \begin{tabular}{|c|c|c|c|c|c|c|c|c|}
                \hline
                BAIR & BDD100K & Caltech & KITTI & KTH & MSR & PENN & PUSH & UCF-101 \\
                \hline
                40 & 40 & 14 & 46 & 33 & 17 & 50 & 10 & 50 \\
                \hline
            \end{tabular}
            \label{tab:num-videos-diff-datasets}
        \end{center}
    \end{table*}
    We apply the video prediction models on nine different datasets typically used in their evaluation.
    These include BAIR~\cite{ebert2017self}, PUSH~\cite{finn2016unsupervised}, KTH~\cite{schuldt2004recognizing}, MSR~\cite{msr2016action}, UCF-101~\cite{soomro2012ucf101}, PENN~\cite{zhang2013actemes}, KITTI~\cite{geiger2013vision}, Caltech Pedestrian~\cite{dollar2011pedestrian} and BDD100K~\cite{yu2018bdd100k}.
    Among the above datasets, the BAIR robot push dataset is highly stochastic i.e.\ the movement of the robotic arm given the current frame is random.
    The other datasets have relatively lower stochasticity, as argued in~\cite{lee2018stochastic}.
    For the sake of simplicity, we refer to these datasets as deterministic datasets.
    Table~\ref{tab:num-videos-diff-datasets} shows the number of videos taken from each dataset.

    \textit{Video Prediction Models}:
    We use a total of seven video prediction models.
    The models can be broadly classified as deterministic and stochastic.
    The deterministic models are trained to predict the future frames, exactly as in the ground truth video.
    The deterministic models we use are PredNet~\cite{lotter2017deep}, MCnet~\cite{villegas2017mcnet}, Future GAN~\cite{aigner2019futuregan} and DYAN~\cite{liu2018dyan}.
    On the other hand, the stochastic models model uncertainty by being trained to predict a distribution of possible futures using noise as an additional input.
    We use videos generated by SAVP~\cite{lee2018stochastic}, SV2P~\cite{babaeizadeh2018stochastic}, SVG-LP~\cite{denton2018stochastic} and some of their ablation models in our database.
    Along with the videos predicted by these models, we also include ground truth or natural videos (videos that are not predicted) from these datasets in our database.
    This helps us validate various aspects of the study, such as biases due to different resolutions and whether the subjects are able to perceive the distortions and accordingly judge the quality of the predicted videos.

    \textit{Distortions}:
    While loss of quality can occur in multiple ways, we broadly observe four different types of distortions in the predicted videos.
    The loss of quality is primarily seen in the form of blurred frames or distorted object shapes.
    The use of pixel level loss measures such as mean squared error in training video prediction algorithms can lead to blurred frames~\cite{lotter2017deep}, as shown in Fig.~\ref{fig:diff-distortions}a.
    We observe that algorithms trained using adversarial loss functions~\cite{villegas2017mcnet,aigner2019futuregan}, result in distortions of object shapes in frames further into the future, as shown in Fig.~\ref{fig:diff-distortions}b.
    This primarily occurs in objects with reasonable motion.
    We also notice the sudden appearance or disappearance of objects, as shown in Fig.~\ref{fig:diff-distortions}c.
    Occasionally, we observe inexplicable color variations during the video trajectory that look unnatural, as shown in Fig.~\ref{fig:diff-distortions}d.

\begin{figure}
        \includegraphics[width=\linewidth]{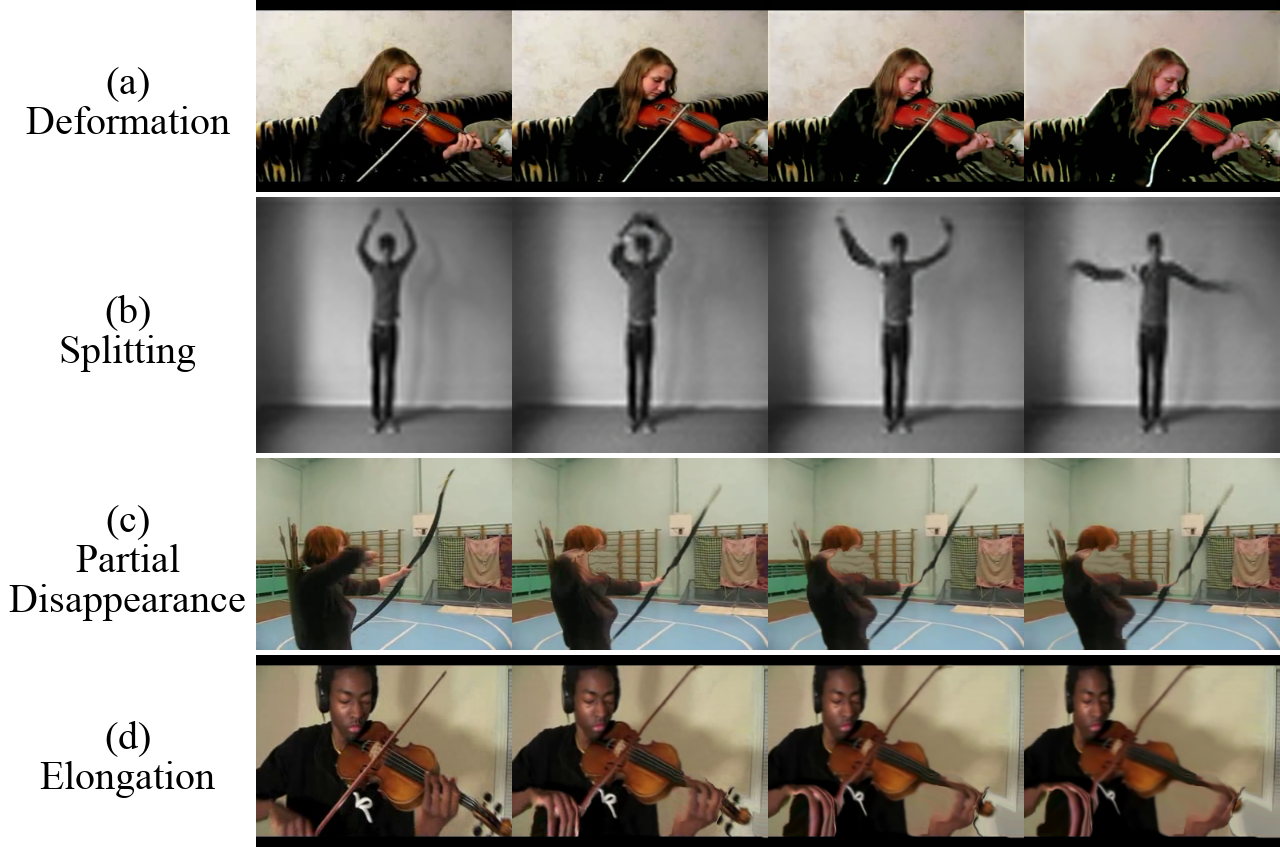}
        \caption{Different kinds of shape distortions observed in predicted videos.
        The sequence of images in each row correspond to the frames of a video.
        Starting from the first frame, we show every fifth frame.
        The first frame is a context frame and the next 3 frames are predicted frames.
        The videos can be viewed on our project website.}
        \label{fig:diff-shape-distortions}
    \end{figure}

    Further, we see different kinds of shape distortions such as deformations (Fig.~\ref{fig:diff-shape-distortions}a), splitting (Fig.~\ref{fig:diff-shape-distortions}b) and elongations of objects (Fig.~\ref{fig:diff-shape-distortions}d).
    In some videos, we witness a combination of shape distortions with object disappearance (Fig.~\ref{fig:diff-shape-distortions}c).
    We note that shape distortions are highly localized, while the rest of the video frame looks completely natural.

    \textit{Selection of Videos}:
    The videos in our database include those generated by applying stochastic models on stochastic datasets, stochastic models on deterministic datasets, and deterministic models on deterministic datasets.
    Using the above combinations, we generate a large number of videos.
    We group these predicted videos based on the type of distortions, such as blur, shape distortion, disappearance and color change, as discussed earlier.
    Among videos that suffer from blur and shape distortions, we manually tag them as low, medium and high quality videos.
    We then roughly select an equal number of blur and shape distorted videos at varying levels of quality, ensuring diversity among the different datasets used to generate the predicted videos.
    We observed that blur and shape distortions were most commonly seen in the predicted videos and they form a significant part of our database.
    Nonetheless, we also select videos with all the other distortions that we observed such as disappearance of objects and color changes, but these videos are fewer in number.
    Thus, we arrive at a total of 300 videos that represent most of the distortions observed in the predicted videos.

    \textit{Video Resolution and Duration}:
    Since different video prediction models available in the literature are trained to generate videos at different resolutions, the videos in our database are of varying resolutions.
    The resolutions include $64 \times 64, 128 \times 128, 160 \times 128$, and $320 \times 240$.
    We discuss the implications of this aspect of the database and the normalization required while conducting the subjective study in Section~\ref{subsec:subjective-study}.
    All videos generated by the prediction algorithms have 4 context frames and 16 predicted frames.
    Following~\cite{lee2018stochastic}, where a small scale subjective evaluation was conducted, we use a frame rate of 4fps for all the videos.
    Thus, each video is of duration 5 seconds during playback.

    \begin{figure}
        \includegraphics[width=\linewidth]{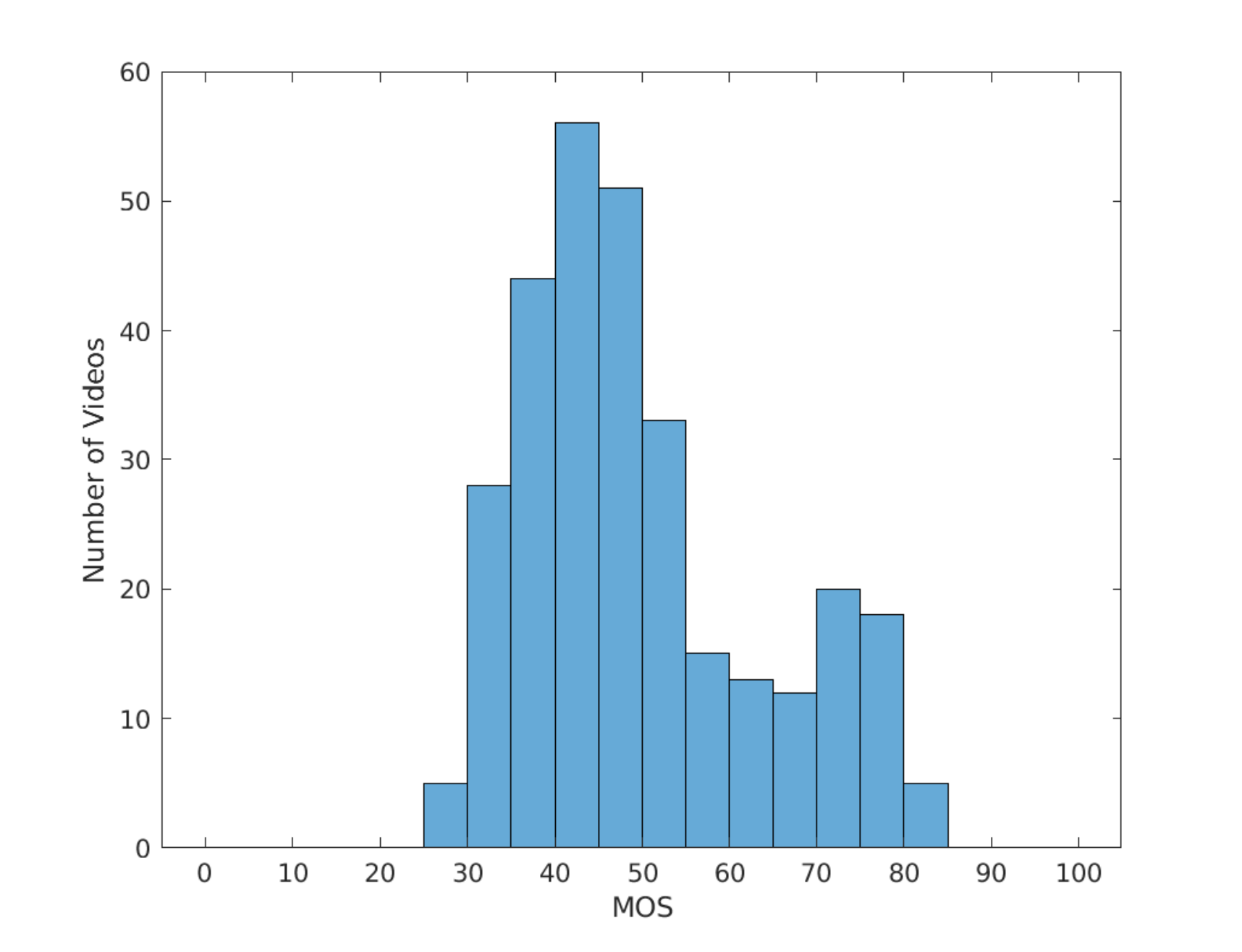}
        \caption{Distribution of Mean Opinion Scores (MOS)}
        \label{fig:mos-distribution}
    \end{figure}

    \subsection{Subjective Study}\label{subsec:subjective-study}
    We conduct a subjective study to assess the quality of the predicted videos.
    Since the subjective evaluation of quality of predicted videos has not been studied before and it is not clear apriori how humans would respond to the task of assessing distortions in predicted videos which are often localized, we conduct the study in a controlled lab environment.
    In our study, 50 subjects participated under calibrated viewing conditions, and all the subjects viewed the videos on a 24 inch LED monitor.
    Most of the subjects were students studying at IISc and hence in the age range 20--30.
    Of the 50 subjects, 35 were male and 15 were female.
    Each subject rated a total of 120 videos, 60 each in two sessions, each session lasting around half an hour and separated by a minimum of 24 hours.
    For each subject, the videos were presented in a random sequence.
    Each video is rated by an equal number of subjects.
    Since there are 300 videos in our database, we obtain a total of 20 human scores for each video.

    Since it is difficult to perceptually understand the lower resolution videos in our database, such videos are upsampled using bicubic interpolation and shown during the subjective study.
    In order to remove any biases in the scoring of such upsampled videos, we employ a double stimulus continuous quality assessment scoring mechanism.
    Here, a natural video with similar content at the same resolution as the evaluation video is also upsampled and shown on the left while the evaluation video is shown on the right.
    Our database broadly consists of four different classes of videos, namely human actions, sports, vehicle driving, and robot pushing videos.
    Every evaluation video belongs to one of these classes.
    A natural video with similar content refers to a natural video from the same class as the evaluation video.
    The subjects are asked to rate the quality of the evaluation video on a scale between 0 and 100, assuming that the natural video shown would correspond to a score of 100.
    We show in Section~\ref{subsec:analysis-of-subjective-scores} that such upsampling does not bias the quality scores of the upsampled videos.

    Since most of the videos in the database show a degradation of quality with time, we asked the subjects to take into account the entire 5s duration video and provide a single holistic score of the video quality.
    The videos are looped continuously and the subjects can view them as long as desired before providing a rating on a continuous scale that appears at the bottom of the screen.
    Every subject is shown 6 videos prior to the start of the study in each session.
    This allows the subject to get a sense of the range of quality levels and different kinds of distortions in the database.

    \textit{Processing of Subjective Scores}:
    We process the collected subjective scores to obtain a mean opinion score (MOS) of quality for every video following well established procedures in VQA~\cite{seshadrinathan2010study}.
    In particular, we subtract the mean and standard deviation of the scores of each subject in each viewing session to obtain `Z-scores'.
    We then apply the subject rejection procedure outlined in ITU-R BT 500.11 recommendation~\cite{itu2002methodology} to remove the outlier subjects.
    In our study, we found 7 out of 50 subjects to be outliers.
    We note that, 6 out of 7 outliers marginally satisfied the rejection criteria.
    The scores from the inlier subjects are then rescaled linearly to lie between 0 and 100, and the MOS for every video is computed as the average Z-score (after rescaling) of every video across all subjects who rated that video.
    Fig.~\ref{fig:mos-distribution} shows the distribution of MOS, where we see that more than 90\% of the scores lie in the range [30, 80].
    Such a distribution of scores presents a challenging test condition for quality assessment methods.
    In Fig.~\ref{fig:mos-distribution}, we observe a small peak around MOS value of 75.
    This peak is due to the presence of natural videos in our database.

    \subsection{Observations from the Subjective Study}\label{subsec:analysis-of-subjective-scores}

    \subsubsection{Consistency of subjects}\label{subsubsec:consistency-of-subjects}
    \begin{figure}
        \includegraphics[width=\linewidth]{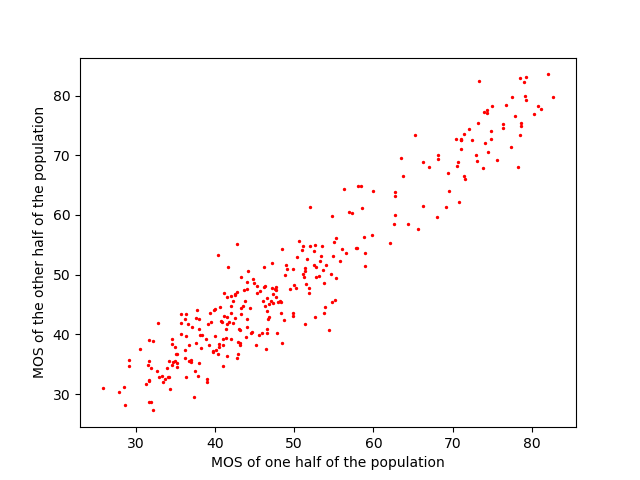}
        \caption{Scatter plot of MOS obtained from two random halves of the population.}
        \label{fig:subject-consistency}
    \end{figure}
    We randomly split the inlier subjects into two halves and compute MOS for each video in each half of the population.
    We then compute the Pearson's linear correlation coefficient (PLCC) between the MOS coming from each half.
    Fig.~\ref{fig:subject-consistency} shows scatter plot of MOS obtained from each half for one such split, where we observe high correlation between MOS from the two halves.
    Further, we compute median PLCC across 100 random splits of the population, which works out to 0.94.
    This shows that the subjects are fairly consistent in assessing the quality of the videos.
    This also provides a reasonable upper bound on the correlation with the subjective scores, which we can expect from objective measures of quality.

    \subsubsection{Validation of our subjective study}
    We now study the average MOS of the natural videos and predicted videos in Table~\ref{tab:analysis-database}.
    We clearly see that the average MOS for natural videos is higher than that of predicted videos.
    This shows that the subjects are able to perceive distortions and rate natural videos with higher scores.

    In order to study the impact of upsampling low resolution videos on the subjective scores, we compare the average MOS of upsampled (for lower resolutions such as $64 \times 64, 128 \times 128, 160 \times 120$) and non-upsampled videos (with higher resolution $320 \times 240$) in Table~\ref{tab:analysis-database}.
    We conduct this test on natural videos presented as test videos to avoid any bias due to the distortions contained in the predicted videos.
    We observe that the average MOS for the upsampled videos is comparable to that of the videos at their original higher resolutions.
    In order to verify the statistical indistinguishability of the MOS in each case, we also conduct t-test~\cite{casella2002statistical} at 99\% significance level.
    The null hypothesis is that the mean of the MOS values for both groups are equal, and the alternate hypothesis is that the means are different.
    The $p$-value of the t-test evaluates to $0.07 (> 0.01)$, and hence the null hypothesis cannot be rejected.
    Thus we conclude that the upsampled videos do not suffer from any biases in their subjective ratings.

    \begin{table}
        \caption{Average MOS for different subsets of videos.
        Standard deviation of the scores and the number of videos in both categories is also shown.
        Note that some videos have both blur and shape distortions and such videos are marked under both categories.}
        \centering
        \begin{tabular}{|Sc|c|c|}
            \hline
            Experiment Type & No. of Videos & Average MOS \\
            \hline
            \rule{0pt}{1ex} \multirowcell{2}{\setstackgap{L}{1.75ex}\Centerstack{Natural Videos\\vs\\Predicted Videos}} & 30 & 76.68 $\pm$ 03.79 \\
            & 270 & 46.97 $\pm$ 10.88 \\
            \hline
            \rule{0pt}{1ex} \multirowcell{2}{\setstackgap{L}{1.75ex}\Centerstack{Upsampled Natural Videos\\vs\\Non-upsampled Natural Videos}} & 16 & 75.50 $\pm$ 3.46 \\
            & 14 & 78.03 $\pm$ 3.68 \\
            \hline
            \rule{0pt}{1ex} \multirowcell{2}{\setstackgap{L}{1.75ex}\Centerstack{Blur\\vs\\Shape Distortion}} & 163 & 45.57 $\pm$ 08.52 \\
            & 200 & 45.32 $\pm$ 10.80 \\
            \hline
            \rule{0pt}{1ex} \multirowcell{2}{\setstackgap{L}{1.75ex}\Centerstack{Stochastic Prediction\\vs\\Deterministic Prediction}} & 73 & 54.26 $\pm$ 12.52 \\
            & 197 & 44.27 $\pm$ 08.78 \\
            \hline
        \end{tabular}
        \label{tab:analysis-database}
    \end{table}

    \subsubsection{How does MOS vary for different distortions?}
    We observe that shape distortions and blur are the two predominant classes of distortions in the predicted videos.
    We roughly classify the videos into those that contain shape distortions and those that contain blur.
    Some videos have both distortions, in which case they are marked under both categories.
    The resulting MOS for the two classes of videos is shown in Table~\ref{tab:analysis-database}.
    We find that the average MOS for videos with blur is roughly equal to the average MOS for videos with shape distortions.
    Among other distortions such as disappearance and color changes of objects, our database has 30 videos with such distortions and their average MOS is 47.41.
    Thus, these distortions appear to be as annoying as the blur and shape distortions discussed earlier.

    \subsubsection{Do stochastic models perform better than deterministic models?}
    As we pointed out earlier, deterministic methods~\cite{mathieu2016deep,villegas2017mcnet} pick only one of the multiple plausible trajectories.
    On the other hand, stochastic approaches train the model to predict multiple future trajectories~\cite{lee2018stochastic,babaeizadeh2018stochastic,denton2018stochastic}.
    Table~\ref{tab:analysis-database} shows the average MOS and standard deviation with respect to the two methods described above.
    We see that the average MOS is lower for deterministic methods when compared to stochastic models.
    We also verify the statistical significance of this observation using t-test~\cite{casella2002statistical} at 99\% significance level.
    The null hypothesis is that the mean MOS scores of the two groups are equal, and the alternate hypothesis is that the mean MOS score of stochastically predicted videos is higher than that of deterministically predicted videos.
    The $p$-value of the t-test evaluates to $6 \times 10^{-9} (< 0.01)$ and hence the null hypothesis can be rejected.
    Thus, we can conclude that the ability of stochastic models to better capture the uncertainty in the future trajectories, allows them to generate videos of higher quality.

    \section{Feature Extraction for Video Quality Assessment}\label{sec:deep-feature-processing}
    We now present two sets of features that are particularly relevant in reliably predicting the quality of predicted videos.
    The first set of features is motivated by the observation that objects in a scene are well represented in the past frames and can be used to measure how representations evolve in future predicted frames.
    Thus we exploit the rich information available in the deep features of objects in the past frames and make motion-compensated comparisons of deep features in predicted frames.
    We capture this idea through motion-compensated cosine similarity based features.
    This feature also helps identify the disappearance or vanishing of objects from the middle of a scene.
    Secondly, we observe that most of the abnormalities in predicted videos occur in regions of motion.
    In order to capture variations in representations in moving regions and also more carefully measure distortions in object shapes, we introduce the notion of rescaled frame differences and compute deep features from such images.
    We note that deep features based on motion compensation and frame differences contain information about higher order concepts such as shape.
    However, they also contain low level vision information related to blur and sharpness as shown through their application in picture quality assessment~\cite{kim2017deep}.
    We provide further details of both features in the following subsections.

    \begin{figure}
        \begin{subfigure}{\linewidth}
            \includegraphics[width=\linewidth]{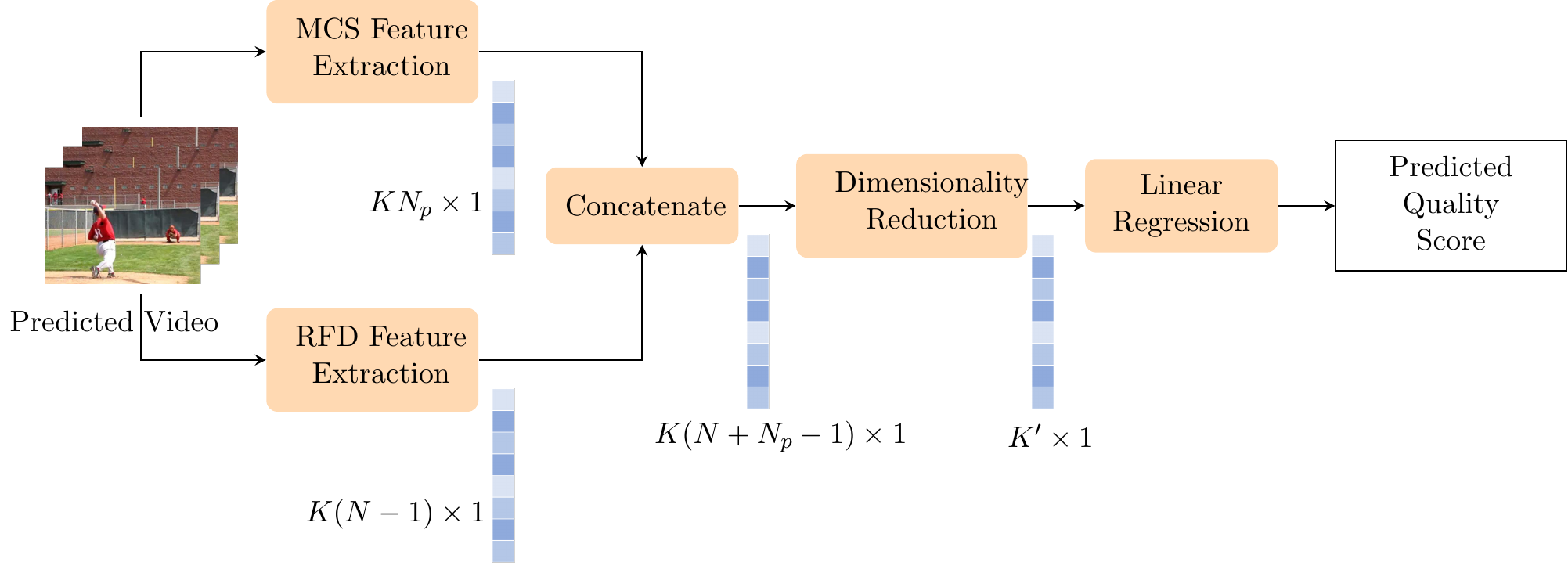}
            \caption{High level architecture of our model. Note that MCS and RFD features are computed on frames and the features across frames are concatenated to obtain video features.}
            \vspace{-2mm}
            \hrulefill
            \vspace{2mm}
        \end{subfigure}
        \begin{subfigure}{\linewidth}
            \includegraphics[width=\linewidth]{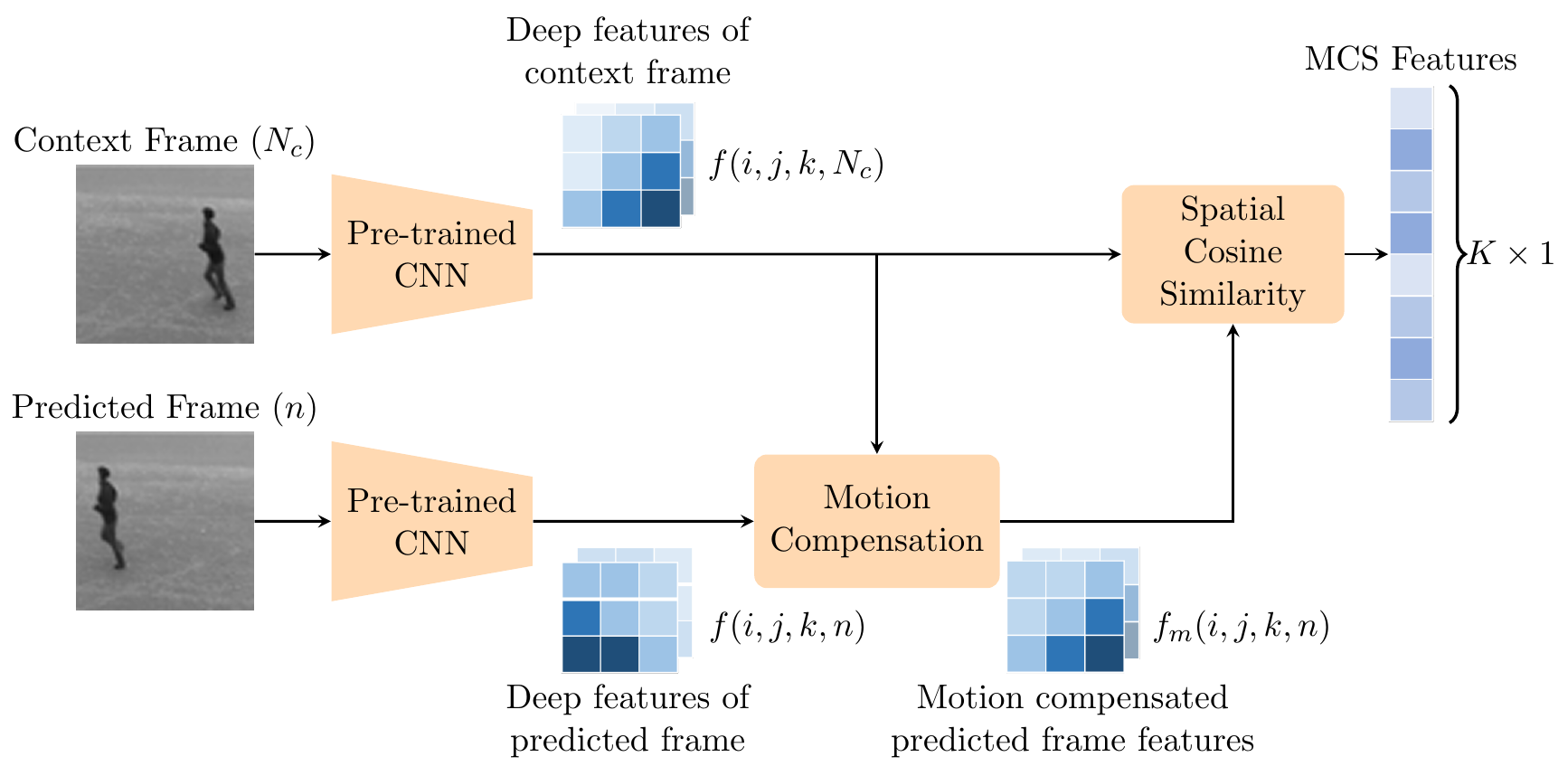}
            \caption{Architecture of the Motion-compensated Cosine Similarity (MCS) feature extraction.}
            \vspace{-2mm}
            \hrulefill
            \vspace{2mm}
        \end{subfigure}
        \begin{subfigure}{\linewidth}
            \includegraphics[width=\linewidth]{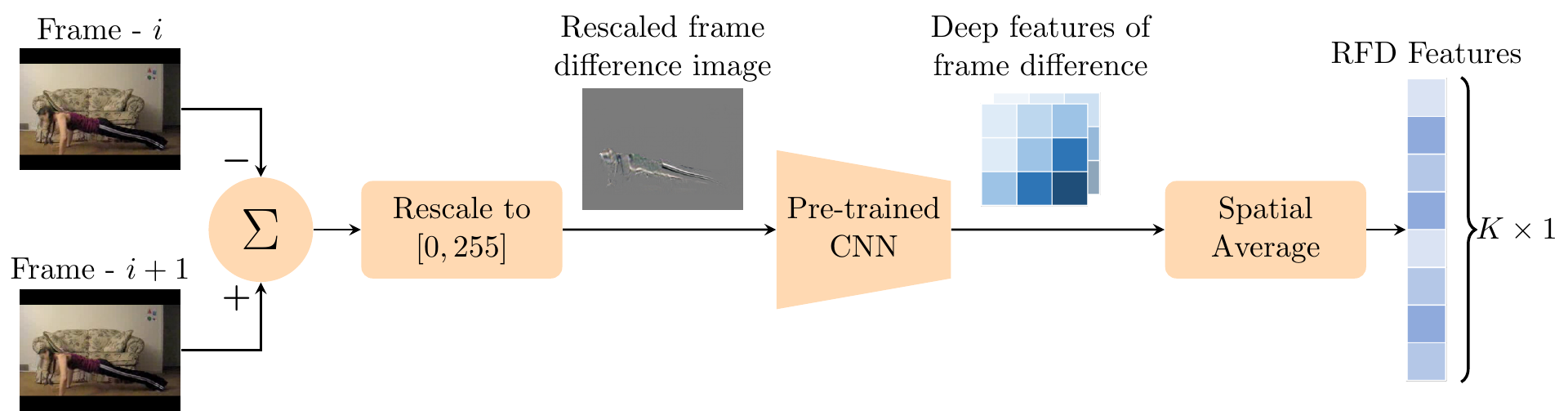}
            \caption{Architecture of the Rescaled Frame Difference (RFD) feature extraction.}
            \vspace{-2mm}
            \hrulefill
        \end{subfigure}
        \caption{Model architecture.}
        \label{fig:model-architecture}
        \vspace{-2mm}
        \hrulefill
    \end{figure}

    \begin{figure*}
        \includegraphics[width=\linewidth]{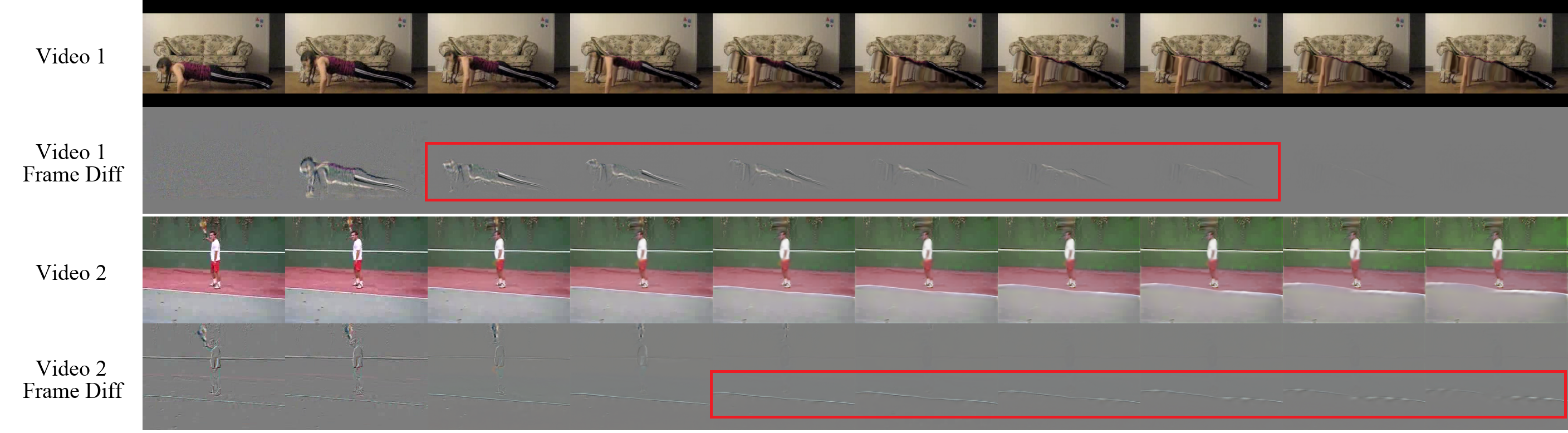}
        \caption{Examples of frame differences for various distortions.
        In the first video, we see the disappearance of the upper torso of the girl.
        In the second video, we observe the movement of the baseline of the tennis court.
        While the first and the last frame may appear largely similar, the movement of the object boundaries is clearly visible in the frame differences.
        The videos can be viewed on our project website.
        }
        \label{fig:frame-diff-examples}
    \end{figure*}

    \subsection{Motion-compensated Cosine Similarity (MCS) features}\label{subsec:mcs-features}
    We illustrate the computation of MCS features in Fig.~\ref{fig:model-architecture}b.
    We experiment with different networks to obtain deep features such as VGG-19, ResNet-50, and Inception-v3 and refer to one such network in the following.
    Let $N$ be the total number of frames, $N_c$ be the number of context frames, and $N_p$ be the number of predicted frames.
    Thus $N = N_c + N_p$.
    Let $K$ be the number of channels in the pre-trained model, at the layer where we tap the features.
    Let $h$ and $w$ be the height and width of the corresponding feature map.

    Let $f(i,j,k,n)$ denote the deep feature at location $(i,j)$ in Channel $k$ in Frame $n$, where $i\in\{1,2,\ldots,h\}$, $j\in\{1,2,\ldots,w\}$, $k\in\{1,2,\ldots,K\}$ and $n\in\{1,2,\ldots,N\}$.
    The cosine similarity between two vectors $\mathbf{p}$ and $\mathbf{q}$ be defined as
    \begin{equation*}
        s(\mathbf{p},\mathbf{q}) = \frac{\mathbf{p}^T\mathbf{q}}{\lVert \mathbf{p} \rVert \lVert \mathbf{q} \rVert}.
    \end{equation*}
    where $\lVert \cdot \rVert$ denotes the two-norm of the vector.
    Let $f(i,j,\cdot,n)$ denote a vector of deep features across channels at location $(i,j)$ in Frame $n$.
    For a given feature $f(i,j,k,N_c)$ in Frame $N_c$ (last context frame), the corresponding motion compensated feature in Frame $n$ with $n>N_c$ is obtained as
    \begin{equation*}
        f_m(i,j,k,n) = f(i',j',k,n),
    \end{equation*}
    where
    \begin{equation*}
        i',j' = \argmax_{i'',j''} s(f(i,j,\cdot,N_c),f(i'',j'',\cdot,n)).
    \end{equation*}
    In other words, for every location in the context frame, we determine the location in the predicted frame with the best cosine similarity in the feature space.
    Thus we obtain the motion compensated features in each predicted frame and compute the MCS feature in Frame $n$ and Channel $k$ as
    \begin{equation*}
        f_{\textrm{MCS}}(k,n) = s(f(\cdot,\cdot,k,N_c),f_m(\cdot,\cdot,k,n)),
    \end{equation*}
    where $f(\cdot,\cdot,k,N_c)$ denotes the vectorized deep features across spatial locations in Frame $N_c$ and Channel $k$ and $f_m(\cdot,\cdot,k,n)$ is also defined similarly.
    This gives us a $K$ dimensional MCS feature vector per frame.
    We concatenate the MCS features from all predicted frames to get a $K \cdot N_p$ dimensional feature vector per video.

    The MCS features are important in capturing several aspects such as object blur, distortion of shapes, abnormal disappearance of objects from the middle of a scene, and change in object color.
    We believe that the natural disappearance of objects from scenes (such as objects moving out of the field of view) can be distinguished from unnatural ones by observing the trajectory of MCS features across frames.
    However, we observe that the occurrence of such events is relatively less likely owing to the limited future duration over which video prediction occurs.

    \subsection{Rescaled Frame Difference (RFD) features}\label{subsec:rfd-features}
    The second set of features we design is based on our observation that shape distortions are highly localized in regions containing motion.
    While optical flow may be used to determine motion masked frames as in~\cite{mathieu2016deep}, the flow estimates tend to be noisy in predicted videos that contain a variety of artifacts.
    In order to overcome this challenge, we resort to measuring frame differences between adjacent frames to capture moving regions.
    However, instead of using such information to mask frames, we rescale the frame differences in the intensity range [0,255] for each color channel and extract deep features from such images.
    The deep features (from VGG-19, ResNet-50, or Inception-v3) of rescaled frame differences enable robust measurement of shape distortions as argued below.

    In Fig.~\ref{fig:frame-diff-examples}, we show examples of rescaled frame differences of two predicted videos from our database.
    We observe that the rescaled frame differences, simultaneously capture both the moving regions of frames as well as the changing contours of moving objects.
    We believe that the visualization of changing contours of moving objects in RFD adds robustness in the design of features along with MCS\@.
    We note that RFD resemble sketch images~\cite{eitz2012humans} in the manner in which object outlines are visible.
    Motivated by the success of deep ResNet features in sketch recognition applications~\cite{zou2018sketchy}, we extract similar features from RFD\@.
    We also note that deep features pre-trained on natural images have been processed by adding a few more layers and successfully applied in other applications for medical and satellite image processing~\cite{raghu2019transfusion,buslaev2018fully}.
    We spatially average the deep features from each RFD to get a single feature per channel, and then we concatenate the features across all frame differences and channels to get a $K \cdot (N-1)$ length feature vector.

    In order to further understand the relevance of deep features of RFD, we compare them with deep features of frames.
    Note that deep features of frames typically capture aspects such as object texture, shape, color, and so on~\cite{zeiler2014visualizing}.
    However, we observe that in RFD in Fig.~\ref{fig:frame-diff-examples}, color, and other local properties tend to get suppressed.
    Thus, the corresponding deep features are primarily sensitive to the shape of the moving objects.
    In order to study this more carefully, for the videos in Fig.~\ref{fig:frame-diff-examples}, we compare the dissimilarity of spatially averaged deep features of frames and RFD between the first context frame and the last predicted frame.
    For Video 1, we observe that the dissimilarity score (1 - cosine similarity) for RFD features is 0.34, while that of frame features is 0.16.
    For Video 2, the corresponding scores are 0.43 and 0.27 respectively.
    This illustrates that the deep features of RFD are more sensitive to variations in object shapes when compared with the features of the frames themselves.

    \subsection{Learning quality from features}\label{subsec:learning-network}
    We concatenate the MCS and RFD features, which gives a $K \cdot (N+N_p-1)$ length vector.
    Since the resulting feature dimension is much higher than the number of videos in our database, we reduce the dimensionality of the feature vector through principal component analysis.
    We select a subset consisting of $K'$ principal components.
    We employ linear regression to predict the quality scores from the $K'$ dimensional feature vector.
    We also experimented with regression models that are directly trained on the high dimensional feature vector.
    Since we observed similar performance, we present our simpler approach involving the feature vector obtained through PCA with linear regression.

    \section{Experiments}\label{sec:experiments}

    \begin{table*}
        \centering
        \caption{Evaluation of Objective Measures of Quality in terms of SROCC, PLCC and RMSE.
        We show the median performance over 100 trials of train-test split of the database.
        Also shown are the standard deviations in the performance across the trials.}
        \begin{tabular}{|l|c|c|c|}
            \hline
            Metric & SROCC & PLCC & RMSE \\
            \hline
            MSE & 0.4044 $\pm$ 0.11 & 0.6578 $\pm$ 0.08 & 10.2556 $\pm$ 0.86 \\
            SSIM~\cite{wang2004image} & 0.5274 $\pm$ 0.09 & 0.6828 $\pm$ 0.07 & 09.9311 $\pm$ 0.89 \\
            MS-SSIM~\cite{wang2003multiscale} & 0.5174 $\pm$ 0.09 & 0.6548 $\pm$ 0.08 & 10.2474 $\pm$ 0.88 \\
            Gradient Difference~\cite{mathieu2016deep} & 0.4908 $\pm$ 0.10 & 0.6838 $\pm$ 0.07 & 10.8074 $\pm$ 1.04 \\
            VGG-19 MSE~\cite{ledig2017photo} & 0.5364 $\pm$ 0.08 & 0.6403 $\pm$ 0.07 & 11.4350 $\pm$ 0.97 \\
            LPIPS v0.1 (VGG-16)~\cite{zhang2018unreasonable} & 0.6053 $\pm$ 0.09 & 0.7566 $\pm$ 0.06 & 08.9861 $\pm$ 0.72\\
            PieApp~\cite{prashnani2018pieapp} & 0.6112 $\pm$ 0.08 & 0.7513 $\pm$ 0.05 & 08.8769 $\pm$ 0.77\\
            DISTS~\cite{ding2020dists} & 0.6272 $\pm$ 0.09 & 0.7592 $\pm$ 0.06 & 08.8064 $\pm$ 0.67\\
            VGG-19 cosine similarity & 0.6404 $\pm$ 0.08 & 0.7506 $\pm$ 0.06 & 08.9538 $\pm$ 0.72 \\
            ST-MAD~\cite{vu2011spatiotemporal} & 0.3730 $\pm$ 0.12 & 0.6516 $\pm$ 0.08 & 10.3446 $\pm$ 0.88 \\
            VMAF~\cite{netflix2020vmaf} & 0.6003 $\pm$ 0.09 & 0.7462 $\pm$ 0.06 & 09.3609 $\pm$ 0.73\\
            \hline
            BRISQUE~\cite{mittal2012no} & 0.0905 $\pm$ 0.11 & 0.0942 $\pm$ 0.11 & 13.8893 $\pm$ 1.27 \\
            NIQE~\cite{mittal2013making} & 0.0819 $\pm$ 0.12 & 0.0698 $\pm$ 0.12 & 15.6844 $\pm$ 1.09 \\
            Inception Score (Entropy of Conditional only) & 0.0828 $\pm$ 0.11 & 0.0458 $\pm$ 0.10 & 15.4043 $\pm$ 1.22 \\
            Video BLIINDS~\cite{saad2014blind} & 0.4072 $\pm$ 0.10 & 0.6200 $\pm$ 0.10 & 12.4202 $\pm$ 1.14 \\
            NSTSS~\cite{dendi2020no} & 0.5798 $\pm$ 0.09 & 0.5900 $\pm$ 0.09 & 11.3086 $\pm$ 1.31 \\
            TLVQM~\cite{korhonen2019two} & 0.6028 $\pm$ 0.08 & 0.6442 $\pm$ 0.07 & 10.3313 $\pm$ 0.97 \\
            VSFA~\cite{li2019quality} & 0.6371 $\pm$ 0.09 & 0.6504 $\pm$ 0.08 & 10.7497 $\pm$ 1.12 \\
            \hline
            Baseline - SSA features - 3D ConvNet & 0.4180 $\pm$ 0.10 & 0.4570 $\pm$ 0.09 & 12.4736 $\pm$ 1.07 \\
            Baseline - SSA features - ResNet-50 & 0.7272 $\pm$ 0.06 & 0.7496 $\pm$ 0.06 & 09.1168 $\pm$ 0.76 \\
            Our Model - VGG-19 & 0.7352 $\pm$ 0.07 & 0.8023 $\pm$ 0.05 & 08.1906 $\pm$ 0.85 \\
            Our Model - Inception-v3 & 0.7976 $\pm$ 0.06 & 0.8363 $\pm$ 0.04 & 07.5792 $\pm$ 0.87 \\
            Our Model - ResNet-50 & \textbf{0.8268 $\pm$ 0.04} & \textbf{0.8626 $\pm$ 0.03} & \textbf{06.9854 $\pm$ 0.68} \\
            \hline
        \end{tabular}
        \label{tab:evaluation-metrics}
    \end{table*}

    \subsection{Evaluation of Objective Quality Measures}\label{subsec:eval-benchmark-metrics}
    We present the evaluation of various measures of quality, spanning FR and NR image and video QA indices including those that are currently used to evaluate predicted videos, deep features of spatial and spatio-temporal networks, and finally our feature design contributions.
    To the best of our knowledge, there exists no other publicly available database to assess the quality of predicted videos.
    Also, the conditions for application of our model such as the need for context frames for video prediction, are quite different from classical VQA.
    Thus, we perform all our experiments on the IISc-PVQA database.

    \subsubsection{Existing measures}
    Several QA indices are popularly used to evaluate predicted videos.
    Among FR image QA metrics, we evaluate MSE, SSIM~\cite{wang2004image}, MS-SSIM~\cite{wang2003multiscale}, gradient difference~\cite{mathieu2016deep}, LPIPS~\cite{zhang2018unreasonable}, PieApp~\cite{prashnani2018pieapp} and DISTS~\cite{ding2020dists}.
    We note that the gradient difference measure is related to a sharpness measure as shown in ~\cite{mathieu2016deep}.
    We also evaluate MSE and cosine similarity in the VGG feature space~\cite{lee2018stochastic,kumar2019videoflow} by tapping the features from the fourth convolutional layer of the fifth block of the VGG-19 network~\cite{ledig2017photo}.
    Among NR image QA indices, we evaluate BRISQUE~\cite{mittal2012no} and NIQE~\cite{mittal2013making} by computing them on each frame and taking their average.
    We also evaluate a modified version of Inception Score~\cite{salimans2016improved} on individual frames by computing the entropy of the conditional distribution alone as a measure of the quality.
    Among video QA measures, we evaluate FR measures such as ST-MAD~\cite{vu2011spatiotemporal} and VMAF v1.5.1~\cite{netflix2020vmaf} and NR indices such as Video BLIINDS~\cite{saad2014blind}, NSTSS~\cite{dendi2020no}, TLVQM~\cite{korhonen2019two} and VSFA~\cite{li2019quality}.
    We train VMAF and the NR measures on our database for a fair comparison.

    \subsubsection{Quality assessment using deep features}
    We present a simple baseline by processing the features extracted from ResNet-50~\cite{he2016deep} model, pre-trained on the ImageNet-1k~\cite{russakovsky2015imagenet} image classification database.
    We tap the features before the global pooling operation, apply simple spatial averaging (SSA) to get a feature vector of dimension $K=2048$ per frame.
    We then concatenate the features from each frame and feed them to a regression model, similar to ours in Section~\ref{subsec:learning-network}.

    Additionally, we present another baseline, using features from the pre-trained 3D ConvNet (C3D) model~\cite{tran2015learning}, successfully used in action recognition on videos.
    We resize the input frames to a resolution of $112 \times 112$, tap spatio-temporal features before the last pooling layer, and process them using PCA and linear regression as the other models.
    The number of principal components is set to 40, based on a simple grid search.
    While ResNet-50 is trained on images, C3D is directly trained on videos.

    \subsubsection{Our model}
    We evaluate our model for quality assessment of predicted videos based on MCS and RFD features using different networks such as VGG-19~\cite{simonyan2015very}, ResNet-50~\cite{he2016deep} and Inception-v3~\cite{szegedy2016rethinking}, all of which are pre-trained on the ImageNet-1k~\cite{russakovsky2015imagenet} image classification database.
    We tap features from the last convolutional layer before the FC layers.
    This results in a choice of $K = 512, 2048, 2048$ for VGG-19, ResNet-50 and Inception-v3 networks respectively.
    For PCA, we set the number of principal components, $K'=240$, based on the number of videos in the training set.
    We also demonstrate the variation in performance with respect to $K'$ in Section~\ref{subsubsec: diff-num-principal-components}.


    \subsubsection{Performance Evaluation}
    We evaluate the different quality assessment indices using Spearman Rank Order Correlation Coefficient (SROCC), Pearson linear correlation coefficient (PLCC), and root mean squared error (RMSE) popularly used in the QA literature~\cite{seshadrinathan2010study}.
    In order to evaluate PLCC and RMSE, a non-linear function is fitted to predict the MOS from the objective scores for objective measures that are not trained on our database.
    All the results are obtained by splitting the dataset into training and testing in the ratio 80:20 over 100 iterations and computing the median performance.
    For models that require dimensionality reduction, the principal components are determined on the respective training sets.
    For a fair comparison of measures that require no training on our database, we evaluate the performance measures in the corresponding test sets of each iteration.

    \subsubsection{Results}
    The results of our experiments are presented in Table~\ref{tab:evaluation-metrics}.
    We only show the magnitude of PLCC and SROCC in the table.
    We see that among the FR measures, VGG-19 cosine similarity achieves the best performance in terms of correlation with the subjective scores.
    We believe that the normalization implicit in the computation of the cosine similarity makes it perform better than VGG-19 MSE\@.
    We notice similar performance of SSIM and MS-SSIM measures, perhaps due to the lower resolution of videos in our database.
    NR image QA indices and Inception Score seem to correlate poorly with human perception while video QA indices perform better than these indices.

    On the other hand, deep features of pre-trained networks extracted from video frames tend to achieve better performance.
    In particular, they outperform NR video QA indices such as Video BLIINDS, NSTSS, and TLVQM, which are also trained on our database.
    We believe that the superior performance of deep features over other QA methods is due to their ability to extract high level features.
    We note that the poor performance of the Conv 3D model may be attributed to the training of this model on action recognition.
    Thus, the resulting features may not capture the spatial distortions in video frames.
    Finally, we observe that our model based on MCS and RFD features performs significantly better than all other measures.
    We see an improved performance in terms of all evaluation measures.
    The lower standard deviation across splits in the performance numbers when compared to other methods also suggests that our model consistently achieves excellent performance across splits.

    \subsection{Ablations and Extended Experiments}\label{subsec:ablations}

    \begin{figure}
        \includegraphics[width=\linewidth]{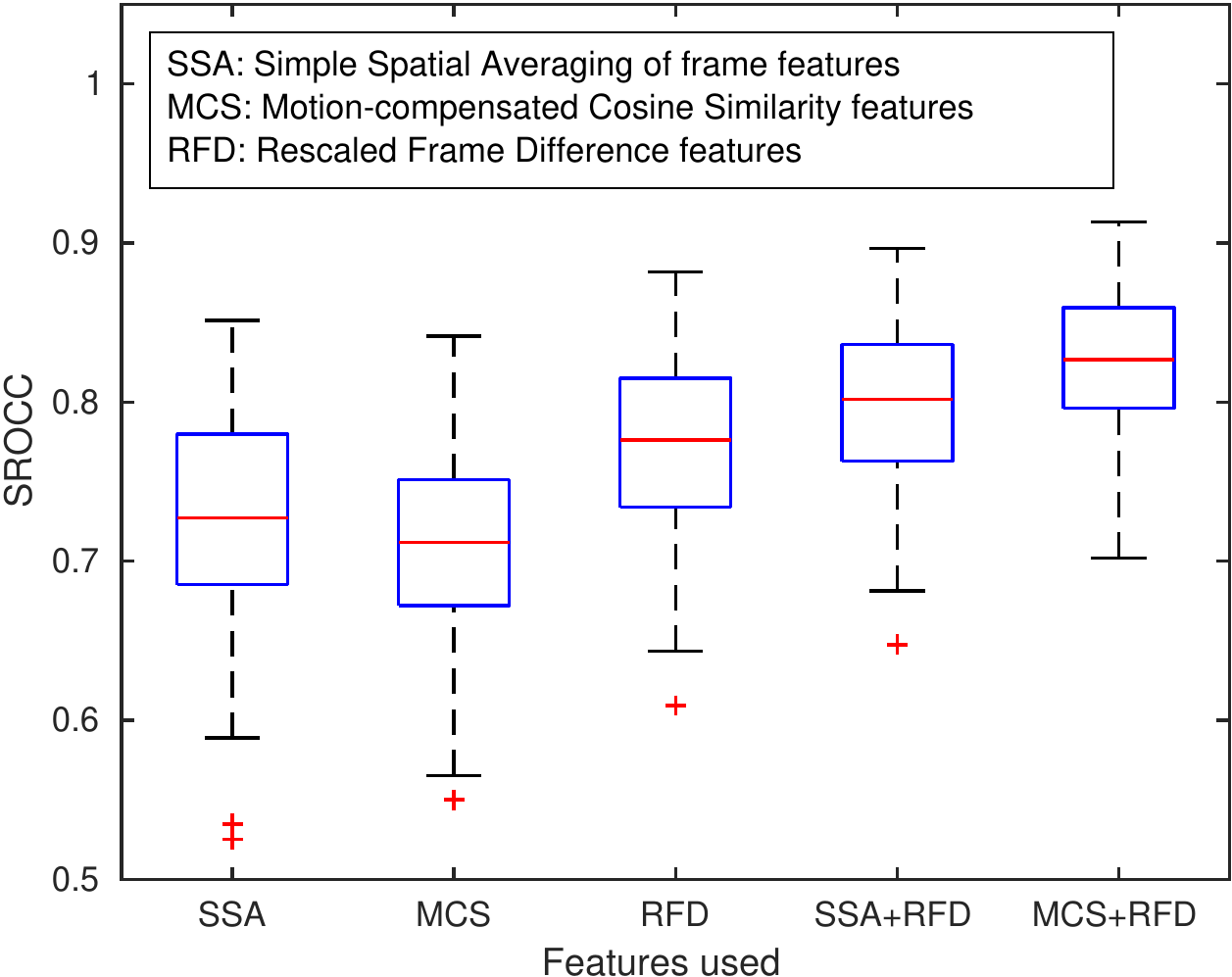}
        \caption{Evaluation of ablation models.
        ResNet-50 features are used for all variants.
        }
        \label{fig:evaluation-ablations}
    \end{figure}

    \begin{figure}
        \includegraphics[width=\linewidth]{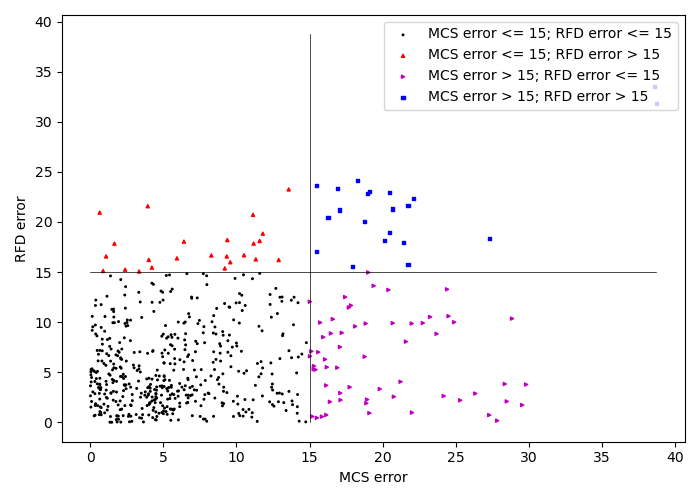}
        \caption{Scatter plot of absolute values of quality score prediction errors by MCS and RFD features.}
        \label{fig:mcs-rfd-complementarity}
    \end{figure}

    \subsubsection{Contribution of individual components}
    Since our model involves two components, the MCS and RFD features, we study the impact of each of the components in Fig.~\ref{fig:evaluation-ablations}.
    We perform this experiment on our model trained on ResNet-50 features, which achieved the best performance.
    We note that RFD features perform better than frame features.
    Further, we see that the combination of the MCS and RFD features leads to a significant improvement in the performance.
    Even though the performance of SSA features is comparable to that of MCS features, MCS features in combination with RFD features perform better than SSA features combined with RFD features.
    Thus, to understand the complementary nature of different sets of features, we quantitatively analyze their dependence.
    In particular, we compute the distance correlation~\cite{szekely2007measuring} of SSA features with RFD features and compare it with the distance correlation of MCS features with RFD features.
    Distance correlation measures the dependence between random vectors (or features in our case).
    This is obtained by first computing pairwise Euclidean distances of samples belonging to each of the two feature types.
    A correlation coefficient is then computed between the resulting two sets of distances.
    We observed a distance correlation of 0.71 between the SSA and RFD features, while the correlation reduces to 0.57 for MCS and RFD features.
    This probably explains why the combination of MCS and RFD features performs better than the combination of SSA and RFD features.

    Further, we regress the MCS and RFD features individually against MOS and visualize a scatter plot of absolute errors in quality score prediction in Fig.~\ref{fig:mcs-rfd-complementarity}.
    The samples correspond to test samples from 10 splits chosen at random.
    We partition the samples into four quadrants based on whether the absolute value of the prediction errors by the two features are above or below a threshold, $15$.
    We observe a significant number of samples for which one of the MCS or RFD features makes a larger error in quality score prediction, but the other feature has a lower prediction error.
    Thus, when used in conjunction, the RFD features can help predict quality better for the samples where MCS features make a larger error and vice-versa.
    We believe that this complementarity between the MCS and RFD features helps our model select the best of both and achieve superior performance.

    \subsubsection{Performance on stochastic videos}\label{subsubsec:perf-stochastic}
    \begin{table}
        \centering
        \caption{Evaluation of Objective Measures of Quality on stochastically predicted videos.
        Only SROCC values are quoted.
        }
        \begin{tabular}{|l|c|c|}
            \hline
            Metric & SROCC \\
            \hline
            VGG-19 cosine similarity & 0.4549 \\
            VMAF & 0.3758 \\
            \hline
            Video BLIINDS & 0.6484 \\
            VSFA~\cite{li2019quality} & 0.7165 \\
            \hline
            Baseline (SSA) - ResNet-50 & 0.7011 \\
            Our Model - ResNet-50 & \textbf{0.7714} \\
            \hline
        \end{tabular}
        \label{tab:eval-stochastic-videos}
    \end{table}

    \begin{figure*}
        \includegraphics[width=\linewidth]{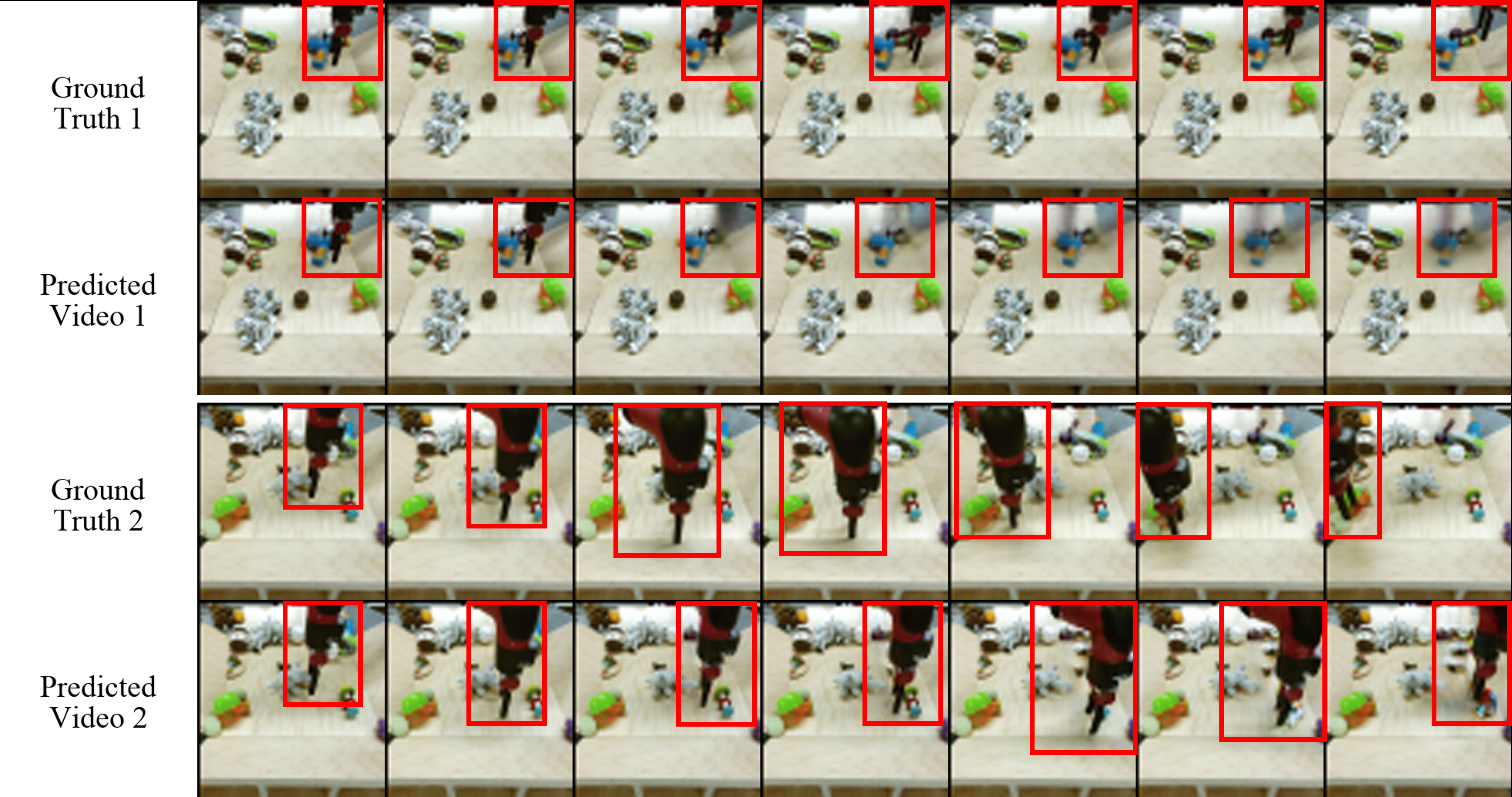}
        \caption{This figure highlights the shortcomings of full reference measures (Section~\ref{subsubsec:perf-stochastic}).
        Two examples of ground truth and predicted videos are shown.
        We show every third frame in the video sequence.
        The first two frames correspond to the context and the next five frames are predicted.
        The red boxes highlight the position of the moving robotic arm.
        The scores of (Predicted Video 1, Predicted Video 2) according to different measures of quality are MSE: (344, 4731), MS-SSIM: (0.9435,0.5586), Cosine Similarity: (0.8860,0.5028), Our Model: (57.40,62.61).
        The corresponding MOS is (46.54,71.65).
        The videos can be viewed on our project website.
        }
        \label{fig:failure-full-reference}
    \end{figure*}
    We now present a couple of examples to support our argument in Section~\ref{sec:introduction} that the inherent stochasticity of the future may reduce the efficiency of full reference measures.
    In Fig.~\ref{fig:failure-full-reference}, we show two examples of ground truth and predicted videos, along with the scores of various full reference measures and our model.
    In Predicted Video 1, we see the disappearance of the robotic arm, which is highly unnatural.
    The movement of the robotic arm in Predicted Video 2 is completely natural and of high quality, just that it is different from Ground Truth 2.
    From the scores shown, we see that all full reference measures fail to capture the quality of videos by indicating that Predicted Video 1 is of higher quality than Predicted Video 2, whereas our model is consistent with human opinion.
    Further, we evaluate various quality measures on stochastically predicted videos of our database in Table~\ref{tab:eval-stochastic-videos}.
    We observe that the performance of the full reference measures is much poorer than the no reference measures.
    Thus we conclude that no reference measures are better equipped to measure quality of predicted videos than full reference measures.

    \subsubsection{Different regression models}\label{subsubsec:different-regression-models}
    \begin{table}
        \centering
        \caption{Performance of our features with different regression models. ResNet-50 features and 240 principal components are used for all variants.}
        \setlength\tabcolsep{5pt} 
        \begin{tabular}{|l|c|c|c|}
            \hline
            Model Variant & SROCC & PLCC & RMSE \\
            \hline
            Linear Regression & \textbf{0.8268} & \textbf{0.8626} & \textbf{6.9854} \\
            Polynomial Regression (degree = 2) & 0.7502 & 0.7599 & 9.5022 \\
            Support Vector Regression (linear kernel) & \textbf{0.8268} & 0.8622 & 6.9906 \\
            Support Vector Regression (rbf kernel) & 0.7135 & 0.7264 & 11.969 \\
            Neural Network & 0.7961 & 0.8547 & 8.3034 \\
            \hline
        \end{tabular}
        \label{tab:diff-regression-models}
    \end{table}

    We also experiment with different regression models such as polynomial regression, Support Vector Regression (SVR) and neural networks.
    For SVR, we found that the linear kernel performs better than radial basis function (rbf) or any other kernel.
    For the neural network, we use two hidden layers with 150 and 50 units respectively.
    We train the model with mean squared error (MSE) loss and Adam optimizer with a learning rate of $0.02$.
    From the results shown in Table~\ref{tab:diff-regression-models}, we note that the performance of linear regression model is slightly superior to that of SVR model, and better than that of polynomial regression model and neural network.

    \subsubsection{Variation in performance with respect to number of principal components}\label{subsubsec: diff-num-principal-components}
    We vary the number of principal components used in our PCA model and show the results of the resulting regression models in Table~\ref{tab:diff-prin-comps}.
    We note that the performance increases with the increase in the number of principal components used.

    
    \begin{table}
        \centering
        \caption{Performance of our model with different number of principal components. ResNet-50 features are used.}
        \setlength\tabcolsep{4.5pt} 
        \begin{tabular}{|l|c|c|c|c|c|c|}
            \hline
            Number of PCA components & 40 & 80 & 120 & 160 & 200 & 240 \\
            \hline
            Median SROCC & 0.79 & 0.81 & 0.82 & 0.82 & 0.82 & 0.83 \\
            \hline
        \end{tabular}
        \label{tab:diff-prin-comps}
    \end{table}

    \subsubsection{Robustness with less training data}\label{subsubsec:less-train-data}
\begin{figure}
        \includegraphics[width=\linewidth]{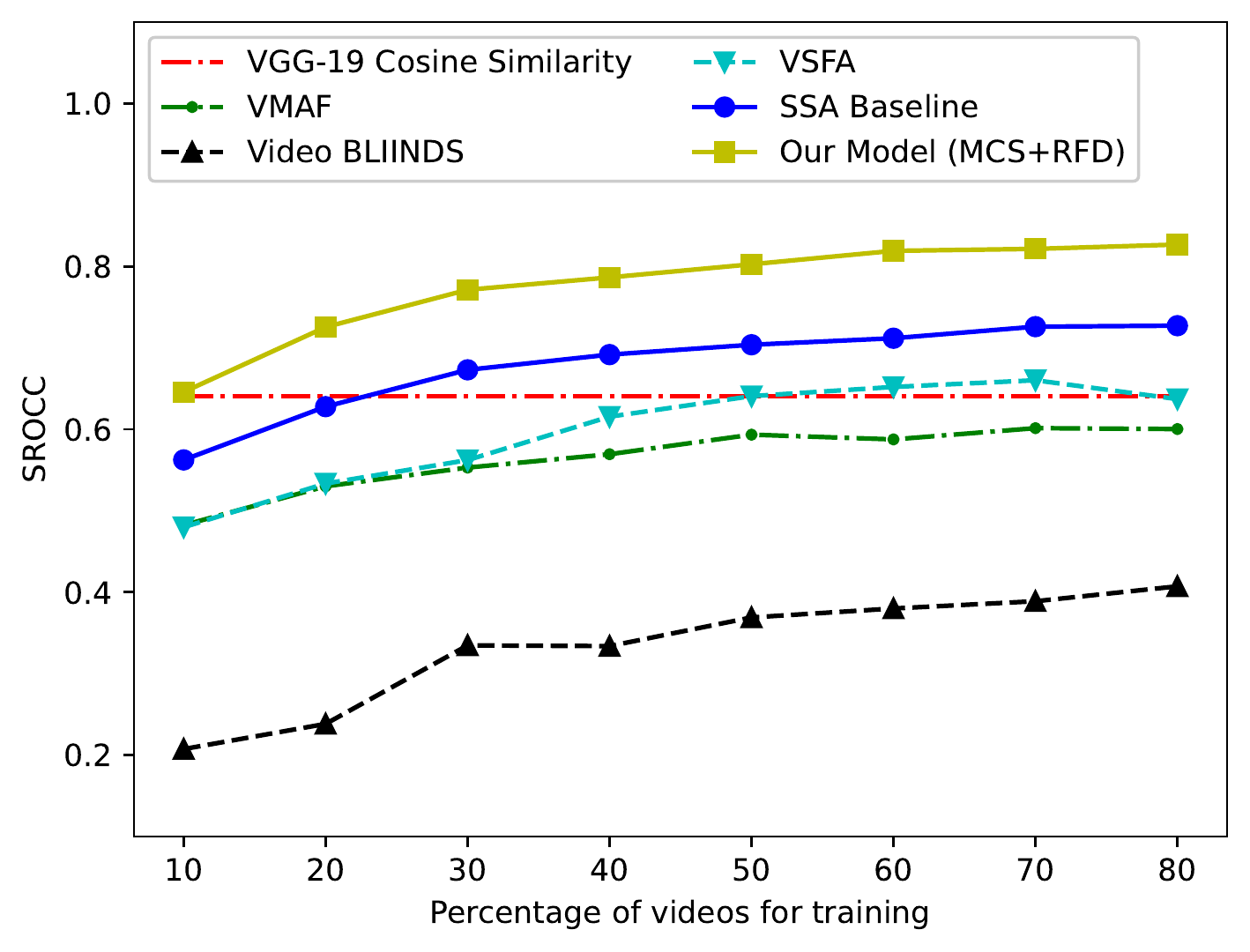}
        \caption{Evaluation of different models for different training set size.
        ResNet-50 features are used for our model.
        Only SROCC values are quoted.
        We observe similar trend w.r.t.\ PLCC and RMSE.
        }
        \label{fig:less-train-data}
    \end{figure}

    We also evaluate the robustness of our model with respect to the amount of training data.
    For a given split of the dataset into training and testing in the ratio 80:20, we build a series of training sets starting with 10\% of the videos and adding 10\% more videos in each step.
    We then evaluate the performance of our model when trained with these subsets as shown in Fig.~\ref{fig:less-train-data}.
    Here, the number of principal components is set to be equal to the number of videos in the respective training set.
    We note that the test data is kept constant across all steps, and in each step, the scores are computed as the median performance across 100 splits.
    For comparison, we also show the performance of other benchmarks and baselines.
    We observe that our model trained with just 10\% of videos in our database, outperforms all existing measures of quality.
    Note that the VGG-19 cosine similarity achieves a constant performance as it is not a training based algorithm.
    Further, we note that our model consistently performs better than other models as the amount of training data increases.
    Thus, we conclude that the robustness of our model with less training data allows for a reliable prediction of quality of predicted videos.

    \subsection{Computational Complexity}\label{subsec:computational-complexity}
    \begin{figure}
        \centering
        \includegraphics[width=\linewidth]{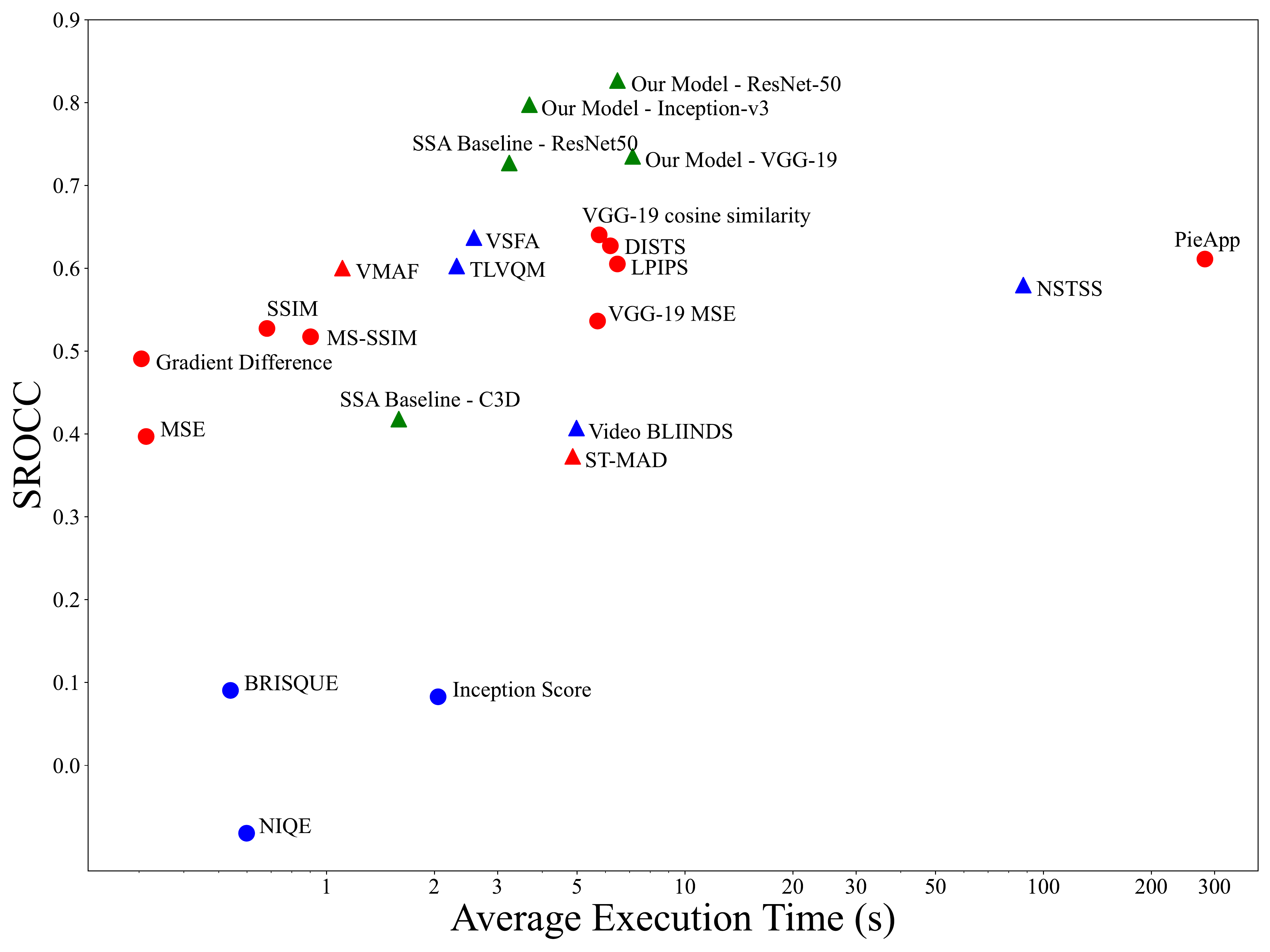}
        \caption{Runtime per video (in seconds) of various QA models is shown against their SROCC scores. The markers in red denote FR measures while those in blue denote NR measures. Circles denote image measures while triangles denote video measures. Finally the markers in green represent our models.}
        \label{fig:computational-complexity}
    \end{figure}
    We compare the computational complexity all models by measuring their testing time per video in the IISc-PVQA database.
    We run all the models on an Ubuntu 64-bit PC with 32 GB RAM and 4GHz octa-core Intel i7-6700K CPU.
    In Fig.~\ref{fig:computational-complexity}, we observe that our models achieve the best SROCC at computational times that are comparable with other deep features based models.
    Although our models involve the computation of motion compensated similarities, since this is evaluated in a feature space at low spatial resolution, it does not appear to have a big impact on runtimes.
    The computational times of our two stream architecture involving MCS and RFD features are comparable to VGG-19 feature based FR measures, which need to compute deep features on both the reference and distorted videos.

    \section{Conclusion}\label{sec:conclusion}
    We build a quality assessment database for video prediction models.
    Our subjective study and benchmarking experiments reveal that existing measures do not correlate well with human perception.
    We show that the MCS and RFD features proposed in this study can capture the quality of predicted videos and outperform all the existing measures of quality.
    We believe that our database will be particularly useful in further research in this area and help design improved models for video prediction.

    Our work in establishing that quality of predicted videos can be assessed reliably by human subjects sets the stage for much larger human studies on more videos, potentially using crowdsourcing.
    We largely focused on predicted videos based on generative models.
    It will be of interest to study the quality of other synthetically generated videos in gaming scenarios.
    Moreover, it will be interesting to understand the role of physics engines in video prediction and quality assessment ~\cite{janner2018reasoning}.
    Finally, we primarily looked at a supervised setting by learning quality from human scores.
    It will also be interesting to explore unsupervised measures of predicted video quality that can be designed by merely having access to a large corpus of natural videos.

    \section*{Acknowledgment}
    The authors would like to thank all the volunteers who took part in the subjective study.

    \ifCLASSOPTIONcaptionsoff
    \newpage
    \fi
    \bibliographystyle{IEEEtran}
    \bibliography{VINE}

\end{document}